\documentstyle[11pt]{article}
\def\thebibliography#1{\section*{REFERENCES\@mkboth
 {REFERENCES}{REFERENCES}}\list
 {\arabic{enumi}.}{\settowidth\labelwidth{[#1]}\leftmargin\labelwidth
 \advance\leftmargin\labelsep
 \usecounter{enumi}}
 \def\newblock{\hskip .11em plus .33em minus .07em}
 \sloppy\clubpenalty4000\widowpenalty4000
 \sfcode`\.=1000\relax}

\newcommand \beq{\begin{eqnarray}}
\newcommand \eeq{\end{eqnarray}}

\catcode`\@=11

\def\ps@myheadings{\let\@mkboth\@gobbletwo
\def\@oddhead{\hbox{} 
\rightmark\hfil\ninerm\thepage}
\def\@oddfoot{}\def\@evenhead{\ninerm\thepage\hfil 
\leftmark\hbox{}}\def\@evenfoot{}
\def\sectionmark##1{}\def\subsectionmark##1{}}

\begin{document}

\begin{titlepage}

\begin{flushright}

{Saclay-T94/068}
\end{flushright}
\vspace*{3cm}
\begin{center}
\baselineskip=13pt
{\Large QCD TRANSPORT THEORY\\}
\vspace{1cm}
 Edmond IANCU\\

{\it Service de Physique Th\'eorique, CEA-Saclay}\\
\baselineskip=12pt
{\it  91191 Gif-sur-Yvette, France}\\
\vskip0.5cm
May 1994
\end{center}

\vskip 2cm \begin{abstract}

Because of the long range of the gauge interactions,
the collective behaviour of quarks and gluons plays a decisive role in the
 transport processes. Collective effects, like Debye screening and Landau
damping, remove the unphysical infrared divergences of the transport
cross-sections and provide finite relaxation rates.
I review here a theory of the plasma collective excitations that has been
 recently developed. It is based on kinetic equations
derived from the general QCD Dyson-Schwinger equations, in the weak coupling
limit. I present new, truly non-abelian, collective excitations, which
correspond to nonlinear color oscillations of the QCD plasma.

\end{abstract}
\vskip 2cm

\begin{flushleft}
Invited talk given at the 5th Conference on the Intersections of Particle
and Nuclear Physics, \\
May 31 -- June 6, 1994, St. Petersburg, Florida, U.S.
\end{flushleft}

\end{titlepage}

\newcommand{\symbolfootnote}{\renewcommand{\thefootnote}
        {\fnsymbol{footnote}}}
\renewcommand{\thefootnote}{\fnsymbol{footnote}}
\newcommand{\alphfootnote}
        {\setcounter{footnote}{0}
         \renewcommand{\thefootnote}{\sevenrm\alph{footnote}}}

\newcounter{sectionc}\newcounter{subsectionc}\newcounter{subsubsectionc}
\renewcommand{\section}[1] {\vspace{0.6cm}\addtocounter{sectionc}{1}
\setcounter{subsectionc}{0}\setcounter{subsubsectionc}{0}\noindent
        {\centerline{\thesectionc. #1}}\par\vspace{0.4cm}}
\renewcommand{\subsection}[1] {\vspace{0.6cm}\addtocounter{subsectionc}{1}
        \setcounter{subsubsectionc}{0}\noindent
        {\it\thesectionc.\thesubsectionc. #1}\par\vspace{0.4cm}}
\renewcommand{\subsubsection}[1]
{\vspace{0.6cm}\addtocounter{subsubsectionc}{1}
        \noindent {\rm\thesectionc.\thesubsectionc.\thesubsubsectionc.
        #1}\par\vspace{0.4cm}}
\newcommand{\nonumsection}[1] {\vspace{0.6cm}\noindent{\bf #1}
        \par\vspace{0.4cm}}

\newcounter{appendixc}
\newcounter{subappendixc}[appendixc]
\newcounter{subsubappendixc}[subappendixc]
\renewcommand{\thesubappendixc}{\Alph{appendixc}.\arabic{subappendixc}}
\renewcommand{\thesubsubappendixc}
        {\Alph{appendixc}.\arabic{subappendixc}.\arabic{subsubappendixc}}

\renewcommand{\appendix}[1] {\vspace{0.6cm}
        \refstepcounter{appendixc}
        \setcounter{figure}{0}
        \setcounter{table}{0}
        \setcounter{equation}{0}
        \renewcommand{\thefigure}{\Alph{appendixc}.\arabic{figure}}
        \renewcommand{\thetable}{\Alph{appendixc}.\arabic{table}}
        \renewcommand{\theappendixc}{\Alph{appendixc}}
        \renewcommand{\theequation}{\Alph{appendixc}.\arabic{equation}}
        \noindent{\bf Appendix \theappendixc #1}\par\vspace{0.4cm}}
\newcommand{\subappendix}[1] {\vspace{0.6cm}
        \refstepcounter{subappendixc}
        \noindent{\bf Appendix \thesubappendixc. #1}\par\vspace{0.4cm}}
\newcommand{\subsubappendix}[1] {\vspace{0.6cm}
        \refstepcounter{subsubappendixc}
        \noindent{\it Appendix \thesubsubappendixc. #1}
        \par\vspace{0.4cm}}

\def\abstracts#1{{
        \centering{\begin{minipage}{30pc}\tenrm\baselineskip=12pt\noindent
        \centerline{\tenbf Abstract}\vspace{0.3cm}
        \parindent=0pt #1
        \end{minipage} }\par}}

\newcommand{\bibit}{\it}
\newcommand{\bibbf}{\bf}
\renewenvironment{thebibliography}[1]
        {\begin{list}{\arabic{enumi}.}
        {\usecounter{enumi}\setlength{\parsep}{0pt}
\setlength{\leftmargin 1.25cm}{\rightmargin 0pt}
         \setlength{\itemsep}{0pt} \settowidth
        {\labelwidth}{#1.}\sloppy}}{\end{list}}

\topsep=0in\parsep=0in\itemsep=0in
\parindent=1.5pc

\newcounter{itemlistc}
\newcounter{romanlistc}
\newcounter{alphlistc}
\newcounter{arabiclistc}
\newenvironment{itemlist}
        {\setcounter{itemlistc}{0}
         \begin{list}{$\bullet$}
        {\usecounter{itemlistc}
         \setlength{\parsep}{0pt}
         \setlength{\itemsep}{0pt}}}{\end{list}}

\newenvironment{romanlist}
        {\setcounter{romanlistc}{0}
         \begin{list}{$($\roman{romanlistc}$)$}
        {\usecounter{romanlistc}
         \setlength{\parsep}{0pt}
         \setlength{\itemsep}{0pt}}}{\end{list}}

\newenvironment{alphlist}
        {\setcounter{alphlistc}{0}
         \begin{list}{$($\alph{alphlistc}$)$}
        {\usecounter{alphlistc}
         \setlength{\parsep}{0pt}
         \setlength{\itemsep}{0pt}}}{\end{list}}

\newenvironment{arabiclist}
        {\setcounter{arabiclistc}{0}
         \begin{list}{\arabic{arabiclistc}}
        {\usecounter{arabiclistc}
         \setlength{\parsep}{0pt}
         \setlength{\itemsep}{0pt}}}{\end{list}}

\newcommand{\fcaption}[1]{
        \refstepcounter{figure}
        \setbox\@tempboxa = \hbox{\tenrm Fig.~\thefigure. #1}
        \ifdim \wd\@tempboxa > 6in
           {\begin{center}
        \parbox{6in}{\tenrm\baselineskip=12pt Fig.~\thefigure. #1 }
            \end{center}}
        \else
             {\begin{center}
             {\tenrm Fig.~\thefigure. #1}
              \end{center}}
        \fi}

\newcommand{\tcaption}[1]{
        \refstepcounter{table}
        \setbox\@tempboxa = \hbox{\tenrm Table~\thetable. #1}
        \ifdim \wd\@tempboxa > 6in
           {\begin{center}
        \parbox{6in}{\tenrm\baselineskip=12pt Table~\thetable. #1 }
            \end{center}}
        \else
             {\begin{center}
             {\tenrm Table~\thetable. #1}
              \end{center}}
        \fi}

\def\@citex[#1]#2{\if@filesw\immediate\write\@auxout
        {\string\citation{#2}}\fi
\def\@citea{}\@cite{\@for\@citeb:=#2\do
        {\@citea\def\@citea{,}\@ifundefined
        {b@\@citeb}{{\bf ?}\@warning
        {Citation `\@citeb' on page \thepage \space undefined}}
        {\csname b@\@citeb\endcsname}}}{#1}}

\newif\if@cghi
\def\cite{\@cghitrue\@ifnextchar [{\@tempswatrue
        \@citex}{\@tempswafalse\@citex[]}}
\def\citelow{\@cghifalse\@ifnextchar [{\@tempswatrue
        \@citex}{\@tempswafalse\@citex[]}}
\def\@cite#1#2{{$\null^{#1}$\if@tempswa\typeout
        {IJCGA warning: optional citation argument
        ignored: `#2'} \fi}}
\newcommand{\citeup}{\cite}

\def\fnm#1{$^{\mbox{\scriptsize #1}}$}
\def\fnt#1#2{\footnotetext{\kern-.3em
        {$^{\mbox{\sevenrm #1}}$}{#2}}}

\font\twelvebf=cmbx10 scaled\magstep 1
\font\twelverm=cmr10 scaled\magstep 1
\font\twelveit=cmti10 scaled\magstep 1
\font\elevenbfit=cmbxti10 scaled\magstephalf
\font\elevenbf=cmbx10 scaled\magstephalf
\font\elevenrm=cmr10 scaled\magstephalf
\font\elevenit=cmti10 scaled\magstephalf
\font\bfit=cmbxti10
\font\tenbf=cmbx10
\font\tenrm=cmr10
\font\tenit=cmti10
\font\ninebf=cmbx9
\font\ninerm=cmr9
\font\nineit=cmti9
\font\eightbf=cmbx8
\font\eightrm=cmr8
\font\eightit=cmti8

\centerline{QCD TRANSPORT THEORY}
\baselineskip=16pt
\vspace{0.4cm}
\centerline{\rm Edmond IANCU}
\baselineskip=13pt
\centerline{\rm Service de Physique Th\'eorique, CEA-Saclay }
\baselineskip=12pt
\centerline{\tenit  91191 Gif-sur-Yvette, France}
\vspace{0.5cm}
\abstracts{      
Because of the long range of the gauge interactions,
the collective behaviour of quarks and gluons plays a decisive role in the
 transport processes. Collective effects, like Debye screening and Landau
damping, remove the unphysical infrared divergences of the transport
cross-sections and provide finite relaxation rates.
I review here a theory of the plasma collective excitations that has been
 recently developed. It is based on kinetic equations
derived from the general QCD Dyson-Schwinger equations, in the weak coupling
limit. 
I present some new, truly non-abelian, collective excitations, which
correspond to nonlinear color oscillations of the QCD plasma.
}
\twelverm   
\baselineskip=14pt

\def\bfepsilon{\mbox{\boldmath$\epsilon$}}
\def\bfgrad{\mbox{\boldmath$\grad$}}
\def\bfgamma{\mbox{\boldmath$\gamma$}}
\def\bfcalA{\mbox{\boldmath${\cal A}$}}
\def\bfcalS{\mbox{\boldmath${\cal S}$}}
\def\bfp{\mbox{\boldmath$p$}}
\def\bfv{\mbox{\boldmath$v$}}
\def\bfj{\mbox{\boldmath$j$}}
\def\bfhp{\mbox{\boldmath$\hat p$}}
\def\bfei{\mbox{\boldmath$e_i$}}
\def\bfe{\mbox{\boldmath$e$}}
\def\bfej{\mbox{\boldmath$e_j$}}
\def\bfk{\mbox{\boldmath$k$}}
\def\bfq{\mbox{\boldmath$q$}}
\def\bfR{\mbox{\boldmath$R$}}
\def\bfC{\mbox{\boldmath$C$}}
\def\bfR{\mbox{\boldmath$R$}}
\def\bfX{\mbox{\boldmath$X$}}
\def\bfx{\mbox{\boldmath$x$}}
\def\bfE{\mbox{\boldmath$E$}}
\def\bfB{\mbox{\boldmath$B$}}
\def\bfy{\mbox{\boldmath$y$}}
\def\bfr{\mbox{\boldmath$r$}}
\def\rmRe{\mbox{\rm$Re$}}
\def\rmIm{\mbox{\rm$ImR$}}


\font\tennbf=cmbx12 \newfam\nbffam
\textfont\nbffam=\tennbf
\def\nbf{\fam\nbffam\tennbf}
\font\tennrm=cmr12 \newfam\nrmfam
\textfont\nrmfam=\tennrm
\def\nrm{\fam\nrmfam\tennrm}

\def\cad{\hbox{c'est \`a dire}}
\def\T{\hbox{temp\'erature}}
\def\Ts{\hbox{temp\'eratures}}
\def\P{\hbox{propri\'et\'e}}
\def\Ps{\hbox{propri\'et\'es}}
\def\E{\hbox{\'equation}}
\def\LE{\hbox{l'\'equation}}
\def\Es{\hbox{\'equations}}
\def\QGP{\hbox{plasma de quarks et de gluons}}
\def\BTD{\hbox{boucles thermiques dures}}
\def\DV{\hbox{d\'eveloppement}}
\def\DNB{\hbox{d\'eveloppement en nombre de boucles}}
\def\AI{\hbox{amplitudes 1-PI}}
\def\UR{\hbox{ultrarelativiste}}
\def\btd{\hbox{boucle thermique dure}}
\def\em{\hbox{\'electromagn\'etique}}

\hyphenation{approxima-tions}
\hyphenation{par-ti-cu-le}
\hyphenation{par-ti-cu-les}

\hyphenation{ac-com-pa-gnees}
\hyphenation{cons-tan-te}
\hyphenation{e-lec-tro-ma-gne-ti-que}
\hyphenation{e-lec-tro-ma-gne-ti-ques}
\hyphenation{im-pe-ra-tif}

\newcommand{\theo}{th\'{e}orie\,\,}
\newcommand{\mod}{mod\`ele\,\,}
\newcommand{\mods}{mod\`eles\,\,}
\newcommand{\theos}{th\'{e}ories\,\,}

\def\square{\hbox{{$\sqcup$}\llap{$\sqcap$}}}   
\def\grad{\nabla}                               
\def\del{\partial}                              

\def\frac#1#2{{#1 \over #2}}
\def\smallfrac#1#2{{\scriptstyle {#1 \over #2}}}
\def\half{\ifinner {\scriptstyle {1 \over 2}}
   \else {1 \over 2} \fi}

\def\bra#1{\langle#1\vert}              
\def\ket#1{\vert#1\rangle}              

\def\simge{\mathrel{%
   \rlap{\raise 0.511ex \hbox{$>$}}{\lower 0.511ex \hbox{$\sim$}}}}
\def\simle{\mathrel{
   \rlap{\raise 0.511ex \hbox{$<$}}{\lower 0.511ex \hbox{$\sim$}}}}


\def\parenbar#1{{\null\!                        
   \mathop#1\limits^{\hbox{\fiverm (--)}}       
   \!\null}}                                    
\def\nunubar{\parenbar{\nu}}
\def\ppbar{\parenbar{p}}


\def\buildchar#1#2#3{{\null\!                   
   \mathop#1\limits^{#2}_{#3}                   
   \!\null}}                                    
\def\overcirc#1{\buildchar{#1}{\circ}{}}


\def\slashchar#1{\setbox0=\hbox{$#1$}           
   \dimen0=\wd0                                 
   \setbox1=\hbox{/} \dimen1=\wd1               
   \ifdim\dimen0>\dimen1                        
      \rlap{\hbox to \dimen0{\hfil/\hfil}}      
      #1                                        
   \else                                        
      \rlap{\hbox to \dimen1{\hfil$#1$\hfil}}   
      /                                         
   \fi}                                         %


\def\subrightarrow#1{
  \setbox0=\hbox{
    $\displaystyle\mathop{}
    \limits_{#1}$}
  \dimen0=\wd0
  \advance \dimen0 by .5em
  \mathrel{
    \mathop{\hbox to \dimen0{\rightarrowfill}}
       \limits_{#1}}}                           

\def\real{\mathop{\rm Re}\nolimits}     
\def\imag{\mathop{\rm Im}\nolimits}     

\def\tr{\mathop{\rm tr}\nolimits}       
\def\Tr{\mathop{\rm Tr}\nolimits}       
\def\Det{\mathop{\rm Det}\nolimits}     

\def\mod{\mathop{\rm mod}\nolimits}     
\def\wrt{\mathop{\rm wrt}\nolimits}     


\def\TeV{{\rm TeV}}                     
\def\GeV{{\rm GeV}}                     
\def\MeV{{\rm MeV}}                     
\def\KeV{{\rm KeV}}                     
\def\eV{{\rm eV}}                       

\def\mb{{\rm mb}}                       
\def\mub{\hbox{$\mu$b}}                 
\def\nb{{\rm nb}}                       
\def\pb{{\rm pb}}                       

%
\def\journal#1#2#3#4{\ {#1}{\bf #2},\  {#4} ({#3})}

\def\AdvPhys{\journal{Adv.\ Phys.}}
\def\AnnPhys{\journal{Ann.\ Phys.}}
\def\EurophysLett{\journal{Europhys.\ Lett.}}
\def\JApplPhys{\journal{J.\ Appl.\ Phys.}}
\def\JMathPhys{\journal{J.\ Math.\ Phys.}}
\def\LettNuovoCimento{\journal{Lett.\ Nuovo Cimento}}
\def\Nature{\journal{Nature}}
\def\NPA{\journal{Nucl.\ Phys.\ {\bf A}}}
\def\NPB{\journal{Nucl.\ Phys.\ {\bf B}}}
\def\NuovoCimento{\journal{Nuovo Cimento}}
\def\Physica{\journal{Physica}}
\def\PLA{\journal{Phys.\ Lett.\ {\bf A}}}
\def\PLB{\journal{Phys.\ Lett.\ {\bf B}}}
\def\PhysRev{\journal{Phys.\ Rev.}}
\def\PRC{\journal{Phys.\ Rev.\ {\bf C}}}
\def\PRD{\journal{Phys.\ Rev.\ {\bf D}}}
\def\PRB{\journal{Phys.\ Rev.\ {\bf B}}}
\def\PRL{\journal{Phys.\ Rev.\ Lett.}}
\def\PhysRept{\journal{Phys.\ Repts.}}
\def\ProcNatlAcadSci{\journal{Proc.\ Natl.\ Acad.\ Sci.}}
\def\ProcRoySoc{\journal{Proc.\ Roy.\ Soc.\ London Ser.\ A}}
\def\RevModPhys{\journal{Rev.\ Mod.\ Phys.}}
\def\Science{\journal{Science}}
\def\SovPhysJETP{\journal{Sov.\ Phys.\ JETP}}
\def\SovPhysJETPLett{\journal{Sov.\ Phys.\ JETP Lett.}}
\def\SovJNuclPhys{\journal{Sov.\ J.\ Nucl.\ Phys.}}
\def\SovPhysDoklady{\journal{Sov.\ Phys.\ Doklady}}
\def\ZPhys{\journal{Z.\ Phys.}}
\def\ZPhysA{\journal{Z.\ Phys.\ A}}
\def\ZPhysB{\journal{Z.\ Phys.\ B}}
\def\ZPhysC{\journal{Z.\ Phys.\ C}}



\setcounter{equation}{0}
\section{COLLECTIVE PHENOMENA AND SCREENING}


Let me start by recalling some typical scales in the high temperature QCD
plasma. In equilibrium, quarks and gluons have generally energies and momenta
of order $T$, and number densities of order $T^3$. Then, the average
interparticle distance $\bar r\sim n^{-1/3} \sim 1/T$ is of the same
order as the  thermal wavelength $\lambda_T=1/k\sim 1/T$ (where
$k\sim T$ is a typical particle momentum), that is, the Pauli principle
cannot be ignored, and quantum distribution functions have to be used.
When  coupled to a slowly varying external perturbation, --- of the type
we consider when looking to transport phenomena ---, the plasma may
acquire a collective behaviour on a typical length/time scale of
order $1/gT$. Collectivity arises since any motion taking place over a
distance scale $\lambda \gg \bar r$ may involve coherently a large number
of particles.

 A typical example is the response of the plasma to a weak
and slowly varying gauge mean field, $A_\mu^a(\omega, {\nbf p})$, with
both $\omega$ and $p$ of order $gT$. In linear response theory, the
induced color current is given by a Kubo-type formula, $j^\mu_a\, = \,
\Pi_{ab}^{\mu\nu}A^b_\nu$, where the (color) polarization tensor
$\Pi_{ab}^{\mu\nu}$ is determined by the motion of the thermal particles
in the presence of the soft field. As it is well known\cite{Silin},
in ordinary electromagnetic plasma the polarization phenomena
 are conveniently described as fluctuations $\delta n({\nbf k}, x)$ of
the distribution functions of the charged particles induced by their
interaction with the average electromagnetic field. For weak fields
and collisionless plasmas,  $\delta n({\nbf k}, x)$ is obtained by solving
the linearized Vlasov equation\cite{Silin}
\beq\label{vlasov}
\frac{\del \delta n}{\del t}\,+\,{\nbf  v}\cdot\frac {\del \delta n}
{\del {\nbf  x}}\,=\, -\,e\,({\nbf  E}+{\nbf  v}\times {\nbf  B})
\cdot\frac{\del n_0}{\del{\nbf  k}},\eeq
where ${\nbf v}$,  ${\nbf k}$, $e$ and $n_0$ denote the particle velocity,
momentum, electric charge and equilibrium distribution function. Once
  $\delta n({\nbf k}, x)$  is known as a functional of the
gauge fields, the induced current follows as
$j^\mu(x) = e\int \frac{d^3k}{(2\pi)^3}\, v^\mu\,
\delta n ({\nbf  k}, x),$ with $v^\mu\equiv (1, {\nbf  v})$,
and the polarization tensor is finally
obtained from $\Pi_{\mu\nu}=\delta j_\mu/\delta A^\nu$. By applying
these formulae to an ultrarelativistic QED plasma, made of
electrons, positrons and photons (with $n_0=1/\bigl(\exp(\beta k)+1\bigr)$,
$\beta \equiv 1/T$, and ${\nbf v} = {\nbf k}/k$), one finds\cite{Silin}
\beq\label{DPi}\Pi_{\mu\nu}(\omega, {\nbf  p})\, =\, m_D^2
\left \{-\delta^0_\mu\delta^0_\nu \,+\,\omega \int\frac{d\Omega}{4\pi}
\frac{v_\mu\, v_\nu} {\omega - {\nbf  v}\cdot {\nbf  p}
+i\epsilon}\right\},\eeq
where the integral $\int d\Omega$ runs over all the directions of
the unit vector ${\nbf v}$ and 
$m_D^2\,\equiv\,
 {e^2 T^2}/{3}$.          
   Of course,
when looking to ultrarelativistic quantum plasmas, one may not trust
the simple kinetic arguments above. However, the field-theoretical one-loop
calculation of $\Pi_{\mu\nu}$ gives, to leading order in $g$, the same result
as eq.~(\ref{DPi}). Quite remarkably, a similar expression (with
$m_D^2\,=\,(2N+N_f)g^2 T^2/6$) is  obtained also for a hot QCD plasma, for
$N$ colors and $N_f$ massless quark flavors\cite{Klimov}.
(In deriving these results, the $g$ dependence of the momentum of
the soft field (recall that $\omega\sim p\sim gT$) plays an essential
role\cite{Pisarski,us}.) Such a `coincidence' suggests that kinetic theory
may be appropriate to understand the polarization properties
of ultrarelativistic plasmas; this is related to the long wavelength
character of the collective motion: $\lambda\sim 1/gT$ is much greater
than both $\lambda_T$ and $\bar r$.

We shall return to QCD kinetic theory in the next section. But, before that,
let me explore the consequences of the polarization tensor (\ref{DPi}) for
the screening of the long range gauge interactions.
The one-gluon (or photon) exchange interaction between two conserved
currents $j_\mu$ and $j_\mu^\prime$ is (with color indices omitted)
\beq\label{V}
V(\omega, {\nbf p})\,=\,\frac{\rho\,\rho^\prime}{p^2 +\Pi_l(\omega, p)}\,+\,
\frac{({\nbf j}\times {\hat {\nbf p}})\cdot ({\nbf j}^\prime
\times {\hat {\nbf p}})}{\omega^2-p^2-\Pi_t(\omega, p)},\eeq
where $ {\hat {\nbf p}}={\nbf p}/p$, $j^\mu=(\rho, {\nbf j})$, etc., and
$\Pi_{l,t}$ denote the longitudinal and transverse components
of $\Pi_{\mu\nu}$, eq.~(\ref{DPi}), as given in Ref. \cite{Klimov}.
In the static limit ($\omega \to 0$), $\Pi_l \to m_D^2$, and the
longitudinal (electric) interaction --- the first term of eq.~(\ref{V}) ---
is screened, with a screening length $\lambda_D=m_D^{-1}\sim 1/gT$.
However, for $\omega =0$, $\Pi_t=0$, and the transverse (magnetic) interaction
of eq.~(\ref{V}) is {\it not} screened. This is similar to what happens in
normal electromagnetic plasmas, where the electric fields are screened
by the mobile charges, while the (static) magnetic fields are not
 (magnetic fields are only screened in a superconductor). For QCD, one
generally expects the generation of a magnetic mass $m_{mag}\sim g^2T$,
via some non perturbative mechanism which remains, as yet, poorly
understood\cite{Pisarski,Pisarski93}. Note, however, that
 independent of the existence
of the magnetic mass, the {\it finite} frequency magnetic fields {\it are}
screened by $\Pi_t$, as firstly observed by
Weldon\cite{Klimov}. At low frequency ($\omega\ll p$), the leading
contribution to $\Pi_t$ is, in fact, given by its imaginary part,
since ${\nrm Im}\,\Pi_t\,\simeq
\,- ({\pi}/4)\,m_D^2\,u\, \propto u,$
   while ${\nrm Re}\,\Pi_t\, \propto u^2$
(with $u\equiv \omega/p$). Therefore, for $\omega\ll p\sim gT$,
the  inverse transverse propagator,
$D_t^{-1}(\omega, p)\equiv \omega^2-p^2-\Pi_t(\omega, p)$,
has the leading behaviour $D_t^{-1}\sim -p^2 + i\pi m_D^2\, \omega/4p$,
and vanishes at the complex wave vector $(i\pi\omega/4 m_D)^{1/3} m_D$;
in coordinate space, this corresponds to field attenuation over a length
scale $\sim (\omega m_D^2)^{-1/3}$, and is analogous to the anomalous skin
effect in a pure metal. Such a dynamical screening is ultimately related
to the Landau damping of  the gauge field, i.e. the
coherent transfer of energy towards the hard
particles which are moving in phase with the field oscillations\cite{Silin}.

\section{FROM FIELD THEORY TO KINETIC EQUATIONS}

The collective motion of the quark-gluon plasma over a length-time scale
$\sim 1/gT$ can be conveniently described
as long wavelength oscillations of gauge and fermionic mean fields to which
the plasma particles couple. The relevant dynamics is described by the QCD
Dyson-Schwinger equations for the $n$-point Green's functions ($n\ge 1$).
I present now the results obtained through a perturbative
 analysis of these  equations in which the {\it leading} terms in an expansion
in powers of $g$ are  consistently preserved\cite{us}.
In doing so, one encounters three types
of approximations, which, in most many-body systems, are
independent approximations, but here are controlled
 by the same small parameter, i.e. $g$. These are the weak coupling
 approximation ($g\ll 1$), the
 long wavelength approximation ($\lambda\sim 1/gT\gg 1/T$), and the small
amplitude approximation (the gauge field strength tensor is limited by
 $F\simle gT^2$, or, equivalently, the gauge potentials satisfy $A\simle T$).
The constraint on the field amplitude ensures the consistency of the
soft covariant derivative $D_\mu =\del_\mu + i g A_\mu^a t^a$ in perturbation
theory: $\del_\mu \sim gT \sim g A_\mu$ for $A_\mu \sim T$. I
emphasize that this does not result in a trivial
linearization of the equations, precisely since the gauge fields
involved in covariant derivatives are to be kept to all orders.

To leading order, the Dyson-Schwinger equations reduce to
 a set of coupled equations for the soft mean fields
and  their induced sources which involve only 2-point functions\cite{us}.
I present here only the equations relevant for the color
oscillations of the plasma, i.e. for the collective motion carrying the gluon
quantum numbers.   The first equation, the
generalization of the Maxwell equation in a polarizable medium,
relates  the gauge mean field $A_\mu^a$ to  the induced color current
 $j^a_\mu$:
\beq\label{ava}
\left [\, D^\nu,\, F_{\nu\mu}(x)\,\right ]^a
\,=\,j_\mu^a(x),
\eeq
where   $F_{\mu\nu}= [D_\mu, D_\nu]/(ig)$.
The induced current  expresses the response of the plasma  particles
to the soft fields $A_\mu^a$ and
 transforms as a color vector in the adjoint representation (see below), so
 that eq.~(\ref{ava}) is gauge covariant.
Both fermions (quarks and antiquarks), and bosons (transverse gluons)
carry color, so that they will all contribute to  the induced current.
To leading order, we have
\beq\label{ji}
j_\mu^{a}(x)\,=\,g\int\frac{d^3k}{(2\pi)^3}\,v_\mu\,
\Bigl\{\,N_{\rm f}\left[ \delta
 n_+^a({\nbf k},x)-\delta n_-^a({\nbf k},x)\right]\,+\,
2N\,\delta N^a({\nbf k},x)\Bigr\},\eeq
where $v^\mu\equiv (1,\,{\nbf  v})$ and ${\nbf  v}\equiv
{\nbf  k}/k$ is the velocity of the hard particle. The
color density matrices 
  $\delta n_\pm \equiv\delta n_\pm^a ({\nbf k},x)\,t^a$
and $\delta N \equiv \delta N^a({\nbf  k},x)\,T^a$ are essentially
 the Wigner transforms of the 2-point correlation fonctions of fermions
and, respectively, transverse gluons, in the presence of the gauge
fields. They are determined
by the following {\it kinetic} equations\cite{us}
\beq\label{n}
\left[ v\cdot D_x,\,\delta n_\pm({{\nbf  k}},x)\right]=\mp\, g\,{\nbf  v}
\cdot{\nbf  E}(x)\, \bigl(dn_0/dk\bigr),\eeq
\beq\label{N}
\left[ v\cdot D_x,\,\delta N({{\nbf  k}},x)\right]=-\, g\,
{\nbf  v}\cdot{\nbf  E}(x)\,\bigl(dN_0/dk\bigr),\eeq
 where $E_a^i\equiv F_a^{i0}$ is the chromoelectric field and
$N_0= 1/(\exp(\beta k)-1)$.
In the abelian case, eq.~(\ref{n}) reduces to the linearized Vlasov
equation discussed in the previous section. In the non abelian case,
the equations above are non linear in the color fields, due to the presence
of covariant derivatives in their left hand side. 

For retarded boundary conditions, such that the fields $A_\mu^a$ vanish
as $x_0\to -\infty$, the induced current given by eqs.~(\ref{ji})--(\ref{N})
has the following expression:
\beq\label{jind}
j^\mu(x)\,=\,m_D^2\int\frac{d\Omega}{4\pi}
\,v^\mu \int_0^\infty d\tau\, U(x,x-v\tau)\, {\nbf  v}\cdot{\nbf  E}
(x-v\tau)\,U(x-v\tau,x),\eeq
where  $U(x,y)$  is the parallel transporter along the straight line
$\gamma$   joining $x$ and $y$,
$U(x,y)=P\exp\{ -ig\int_\gamma dz^\mu A_\mu(z)\}$.
Because of the parallel transporters, $j^\mu$ is non-linear in $A_\mu^a$
to all orders. This is, of course, a consequence of the non abelian
 gauge symmetry: the use of parallel transporters is the only way to
 conciliate the non-locality and the gauge covariance of $j^\mu$.

By successively differentiating eq.~(\ref{jind}) with respect to the
gauge fields, one obtains an infinite series of polarization amplitudes
which characterize the propagation and the mutual interactions of the soft
fields, at leading order. One thus recovers the tensor $\Pi_{\mu\nu}$
of eq.~(\ref{DPi}), and, more generally, all the ``hard thermal loops''
identified in one loop diagrams by Braaten and Pisarski\cite{Pisarski}
and by Frenkel and Taylor\cite{Frenkel}.


The solutions of the field equations (\ref{ava}) with the induced current
 (\ref{jind}) describe long wavelength color waves propagating through
 the plasma. These are  generally
 complicated, integro-differential and non-linear equations.
However, they simplify for  {\it plane wave} solutions, i.e.
for $A_\mu^a(x)={\cal A}_\mu^a(p\cdot x)$, where $p^\mu=(\omega, {\nbf p})$
is a fixed 4-momentum; in this case, $j^\mu$ becomes {\it local}
and {\it linear} in the gauge potentials\cite{pw}:
$j_\mu^a(x)=\Pi_{\mu\nu}(\omega,
{\nbf p}){\cal A}^\nu_a(p\cdot x)$, with $\Pi_{\mu\nu}$ given
by eq.~(\ref{DPi}). Then, the field equations reduce to dynamical
systems\cite{pw}, whose properties can be investigated both numerically and
analytically.
As an example, let me consider transversally polarized
plane waves for the color group $SU(2)$. In the covariant gauge $p^\mu {\cal A}
_\mu^a=0$, a particular such configuration is
${\cal A}^0=0$, ${\bfcalA}(z)\,=\,h_1(z)\,{\nbf e}_1\,T^1\,+
\,h_2(z)\,{\nbf e}_2\,T^2$,               
         where $z\equiv p\cdot x$,  ${\nbf e}_i\cdot {\nbf p}=0$,
${\nbf e}_i\cdot {\nbf e}_j =\delta_{ij}$, and the functions $h_i(z)$
satisfy 
\beq\label{tsys}
(\omega^2 -p^2)\,(\ddot h_1\,+\,\,h_1)\,+\,g^2\,h_2^2\,h_1
&=&0,\nonumber\\
(\omega^2 -p^2)\,(\ddot h_2\,+\,h_2)\,+\,g^2\,h_1^2\,h_2 &=&0,\eeq
where the linear (mass) terms in $h_1$ and $h_2$ arise from the
induced current\cite{pw}. Besides,  $\omega$ and $p$ are related through
the dispersion equation $\omega^2-p^2-\Pi_t(\omega, p)=0$.
Numerical studies of this non-linear system have shown that
its solutions are quasi-periodic for small amplitudes
($h_i\ll T$), when the non-linear terms are not important, but they
become unstable as the amplitude is incresed;
for $h_i\sim T$, the system  presents a chaotic behaviour, whose
physical understanding would be interesting.

Even if the system (\ref{tsys}) is non-integrable,  some
 particular (periodic) solutions to it may  still be found
 analytically\cite{pw}. We have, for instance, the symmetric solution
$h_1 = h_2 = h_0 \,{\nrm cn} (\nu z; \kappa)$, where ${\nrm cn} (z; \kappa)$
is the Jacobi elliptic cosine of modulus $\kappa$, and the constants
$h_0$, $\nu$ and $\kappa$ are related to the energy
density of the plane wave\cite{pw}.

\section{QCD TRANSPORT COEFFICIENTS}

In this section, I review recent calculations of transport coefficients
for a weakly interacting hot QCD plasma. In field theory, the systematic
study of dissipative phenomena requires either the use of Kubo-type
formulae, --- which relate the transport coefficients to various correlation
functions of the energy-momentum tensor\cite{Hosoya} ---, or the use
of quantum transport equations, with collision terms included. So far,
both these approaches are afflicted with serious problems, reflecting the
difficulties in constructing a consistent perturbation theory for hot
gauge theories. For example, one encounters infrared divergences
associated to the long range static magnetic interactions, or to the
massless particles. Besides, the characteristic scales (as $gT$ or
$g^2 T$) involve powers of $g$, so that they interfere in a non trivial
way with the usual expansion in powers of the interaction vertices.
We have seen, in our derivation of collisionless kinetic equations\cite{us},
that it is important to take this properly into account in order to obtain
consistent equations which are gauge covariant. A similar analysis
of the collision terms is still lacking, in spite of several recent
attempts\cite{Elze}.

An alternative, more pragmatical --- and, as yet, also more successful ---
point of view, is to assume the existence of a Boltzmann transport
equation with binary collision terms, as for non relativistic many body
systems with well defined quasiparticles\cite{BP}.
The first approaches\cite{Gavin} employed the relaxation
time approximation, where the collision integral is expressed via the
characteristic time $\tau$ describing the rate at which local equilibrium
is restored --- through collisions --- after an initial perturbation.
If $\sigma_{tr}$ is the specific (transport) cross section, and
 $n$ is the number density
of the scatterers ($n\sim T^3$), then $\tau\sim 1/n\sigma_{tr}$. For processes
which involve energy and momentum dissipation,
 $\sigma_{tr}= \int d\Omega\,(d\sigma/d\Omega)\,(1-\cos \theta)$, where the
$(1-\cos \theta)$ weight arises because large angle scatterings are most
effective in momentum degradation: for small $\theta$, the transferred
momentum $p$ is also small ($p^2\propto (1-\cos \theta)$).

In the Born approximation, the exchange of a {\it bare} gluon
leads to the Rutherford differential cross section, $(d\sigma/d\Omega)\propto
g^4/\sin^4 (\theta/2)\,\propto g^4/p^4$, and the corresponding $\sigma_{tr}$
is logarithmically infrared divergent. However, for $\omega$, $p \simle gT$,
the medium effects are important, as we have seen in Sec. 2, and the
bare matrix element $\sim 1/(\omega^2-p^2)^2$ should be replaced with
the effective interaction (\ref{V}). Then, the various screening mechanisms
reviewed before lead to a {\it finite} transport cross-section. This
is obvious for the electric interaction, since now  $(d\sigma_{el}
/d\Omega)\propto g^4/(p^2+m_D^2)^2$, but it is also true for the magnetic
interaction, as firstly notified by Baym et al.\cite{Baym}:
 $(d\sigma_{mag} /d\Omega)\propto g^4/[p^4+ (\pi \omega m_D^2/4p)^2]$,
and the dynamical screening occuring for $\omega \not = 0$ is sufficient
to insure a finite value for $\sigma_{tr}$: $\sigma_{tr}\sim (g^2/T)^2\,\ln
(T/m_D)$. Then, the typical momentum relaxation time is $\tau_{mom}^{-1}\sim
g^4 T \ln (1/g)$. The various transport coefficients --- as viscosities,
electric conductivity, or flavor diffusion --- are related to $\tau_{mom}$
by standard formulae\cite{Gavin}; more elaborate calculations use variational
methods to solve the Botzmann equation\cite{Baym}.

Contrary to the scattering cross-section, the total interaction rate
 $\sigma= \int d\Omega\,\\(d\sigma/d\Omega)$ remains logarithmically divergent
even when the effective propagator is used for the exchanged gluon.
Accordingly, quantities like the quasiparticle damping
rates\cite{Pisarski,Pisarski93}  are sensitive
to the magnetic scale $g^2 T$. It has been argued
recently\cite{Gyulassy93} that a similar sensitivity also occurs for
the color relaxation times $\tau_{color}$, which measures how rapidly
dissipates a local color excitation. One found\cite{Gyulassy93}
 $\tau_{col}^{-1}\sim g^2 T \ln (m_D/m_{mag})\sim g^2 T \ln (1/g)$,
so that  $\tau_{col}\sim g^2 \tau_{mom}$, and the color diffusion is
smaller by a factor $g^2$ than spin or flavor diffusion. A similar
suppresion appears for the color conductivity. The reason for this behaviour
is that color may be easily exchanged by the gluons in the forward scattering
processes, so that the respective cross section do not pay for the extra
factor $(1-\cos \theta)$.

It would be both interesting and important to analyze QCD transport phenomena
in a more rigorous way, by consistently deriving kinetic equations from quantum
field theory, with attention to the specific scales and to the gauge
symmetry. As suggested by our leading order analysis, the kinetic equations
obtained from truncated Dyson-Schwinger equations may be an
efficient way to study the non perturbative long range behaviour
of the high temperature phase of QCD.


\end{document}